\documentstyle[aps,prl,epsf]{revtex}
\begin{document}
\twocolumn[{\hsize\textwidth\columnwidth\hsize\csname
@twocolumnfalse\endcsname
\title{
\draft
Spatiotemporal Stochastic Resonance in
the Swift-Hohenberg Equation
}
\author{
J. M. G. Vilar and J. M. Rub\'{\i}
}
\address{
Departament de F\'{\i}sica Fonamental, Facultat de
F\'{\i}sica, Universitat de Barcelona, Diagonal 647,
E-08028 Barcelona, Spain 
\date{\today}
}
\maketitle
\widetext
\begin{abstract}
\leftskip 54.8pt
\rightskip 54.8pt

We show the appearance of spatiotemporal  stochastic resonance in
the  Swift-Hohenberg  equation.  This  phenomenon  emerges when a
control   parameter  varies   periodically  in  time  around  the
bifurcation  point.  By using general  scaling  arguments  and by
taking  into   account  the  common   features   occurring  in  a
bifurcation, we outline possible manifestations of the phenomenon
in other pattern-forming systems.

\end{abstract}
}]
\narrowtext

In the last years the  phenomenon of  stochastic  resonance  (SR)
\cite{Benzi,Maki,Maki2,JSP,Moss,Wies,Wiese,thre,mio} has been the
subject  of  intense   activity,   to  the  extent  of   becoming
cross-disciplinary  due to the great  number of  applications  to
different  fields of  science.  The main  result  of SR, which in
some  ways  could  be   considered   counterintuitive,   shows  a
constructive  role of noise, since the  response of a system to a
periodic signal may be enhanced with the addition of an optimized
amount  of  noise.  In this  sense,  the  presence  of SR is well
characterized  by the  appearance  of a  maximum  in  the  output
signal-to-noise ratio (SNR) at a certain nonzero noise level.

In spite of the fact  that  spatial  heterogeneity  is one of the
most common features of systems away from  equilibrium, up to now
there has not been a complete  understanding about the phenomenon
of SR in spatially extended systems.  In this context, only a few
recent    results    are    available    from   the    literature
\cite{spi,array,phi4,wio}.  A   remarkable   feature  of  spatial
systems is their  possibility  to develop  patterns that can form
via bifurcations, for instance from the spatially  uniform state,
as  a  control  parameter  is  varied  \cite{cross,ga}.  In  this
situation,  stochastic  perturbations  play a crucial role in the
initial stages of pattern formation to the extent that the system
may display macroscopic manifestations of thermal noise.

In this Letter we will show that temporal periodic  variations of
the  control  parameter  in the  presence  of noise  can  lead to
spatiotemporal  stochastic  resonance  (STSR)  when the system is
close to a bifurcation  point.  To be explicit, we will treat the
Swift-Hohenberg  (SH)  equation  \cite{cross,swi,ho}  in  detail,
which models  Rayleigh-B\'enard  convection  near the  convective
instability.  This  example,  which  constitutes  the paradigm of
pattern  formation,  exhibits  many  features  common to  systems
around a bifurcation  point.  In this regard, and for the sake of
generality we will outline  applications  of our results to other
situations.

In the  vicinity  of  the  instability  point,  Rayleigh-B\'enard
convection  can  be  described  by  means  of the  stochastic  SH
equation, which in dimensionless spatial units is given by
\begin{equation}
\label{SH}
{\partial \psi \over \partial t}=
h(t)\psi -q(1+\nabla^2)^2\psi -g\psi^3
+\xi(\vec r,t) \;\; .
\end{equation}
Here the control parameter $h(t)=-\kappa+\alpha \sin(\omega_0t)$,
with $\kappa$,  $\alpha$ and $\omega_0$  constants,  reflects the
presence of an external  periodic  forcing, due, for instance, to
variations of the  temperature  difference  between the plates of
the convective  cell  \cite{cross,ho,ga2}.  Moreover, $g$ and $q$
are parameters which depend on the  characteristics of the system
and  $\xi(\vec  r,t)$ is Gaussian  white noise with zero mean and
second moment
$\left<\xi(\vec r,t)\xi(\vec r^\prime,t^\prime)\right>
=2D\delta(\vec r - \vec r^\prime)\delta(t-t^\prime)$,
defining the noise level $D$.

In order to describe  the  temporal  evolution  of the system, we
will  consider the  convective  heat flux, which in this model is
given by
\begin{equation}
\label{J}
J(t)= c \int \psi(\vec r,t)^2 d \vec r\;\; ,
\end{equation}
where $c$ is a constant depending on the physical characteristics
of the system.  This magnitude constitutes the order parameter of
the  transition  from the  homogeneous  state to the state  where
spatial structures  develop.  In regards to spatial order, it can
be revealed by the time-averaged structure factor
\label{S}
\begin{equation}
S(k)=\left<\hat\psi_k\hat\psi_{-k}\right>_t \;\; ,
\end{equation}
where  $\hat\psi_k$ is the spatial Fourier transform of the field
$\psi$ and $<\;>_t$  indicates  time and noise  average.  A sharp
peak in this magnitude  makes the presence of an ordered  spatial
structure manifest.

Since we are  interested  in the  effects of noise, we will first
analyze how a small  amount of noise  affects our system.  In the
absence of noise,  irrespective  of the  initial  condition,  the
field $\psi$ goes to zero at large times.  For a sufficiently low
noise level, and far from the possible initial  transient, $\psi$
is small and the  nonlinearity  in Eq.  (\ref{SH})  does not play
any role.  In this situation, by using dimensional analysis it is
easy to see how the  characteristic  magnitudes scale with noise.
We note that, when the  linearized  equation is  considered,  any
dimensionless parameter cannot depend on the noise level, because
only $D$  involves  the  dimensions  of the  field  $\psi$.  Thus
$\psi$ scales with the noise as $\psi \propto \sqrt{D}$.  The SNR
has dimensions of the inverse of time \cite{mio}, then for $J(t)$
it is given by
\begin{equation}
\label{SNR}
{\rm SNR}=\omega_0f_1\left(\gamma \right) \;\; ,
\end{equation}
where $f_1$ is a dimensionless  function which depends on the set
of dimensionless parameters $\kappa/\omega_0$,  $\alpha/\omega_0$
and $q/\omega_0$, denoted by $\gamma$.  We then conclude that the
SNR does not depend on the noise level.  However, both the signal
and  output  noise  scale  with the  noise as  $D^2$.  This  fact
indicates that the output signal increases when noise  increases.
Additionally,  the  structure  factor also follows a scaling law:
$S(k) \propto D$.

Since  for low noise  level the SNR does not  depend  on $D$, the
lowest order  correction in $D$ to the constant  value of the SNR
comes  from the  nonlinear  term.  To  elucidate  its form, it is
convenient to rewrite Eq.  (\ref{SH}) in the following way
\begin{equation}
\label{SH2}
{\partial \psi \over \partial t}=
[-(\kappa+g\psi^2)+\alpha \sin(\omega_0t)
-q(1+\nabla^2)^2]\psi  +\xi(\vec r,t) \;\; .
\end{equation}
Due to the fact that for low noise level  $\psi^2  \propto D$, in
first  approximation  $\kappa+g\psi^2$  can be interpreted  as an
effective parameter
$\tilde \kappa =  \kappa+g\omega_0^{-1}f_2(\gamma)D$,  with $f_2$
being  a  positive  dimensionless   function.  Consequently,  Eq.
(\ref{SH2}) has the same form as the linearized version for which
the scaling law for the SNR [Eq.  (\ref{SNR})]  has been derived.
By replacing  $\kappa$ by $\tilde \kappa$ in Eq.  (\ref{SNR}) and
expanding around $\kappa$ we then obtain
\begin{equation}
\label{SNR2}
{\rm SNR} \approx \omega_0f_1\left(\gamma \right)
+ { \partial f_1 \over \partial \kappa}
g\omega_0^{-1}f_2(\gamma)D  \;\;
\end{equation}
which includes the lowest order  correction in $D$ to the SNR due
to the nonlinear  term.  An important  consequence of this result
is the fact that the  knowledge of the  dependence  of the SNR on
the parameter  $\kappa$, for the linearized  equation, enables us
to  predict  the  presence  of SR  when  the  nonlinear  term  is
considered.  If $f_1$ is an increasing function of $\kappa$ then
$\partial  f_1 / \partial  \kappa $ is positive and  consequently
the SNR is an  increasing  function  of $D$ for low noise  level.
Since for high $D$ the SNR decreases, one then concludes  that it
exhibits a maximum thus indicating the presence of SR.

\begin{figure}[t]
\centerline{
\epsfxsize=5.5cm 
\epsffile{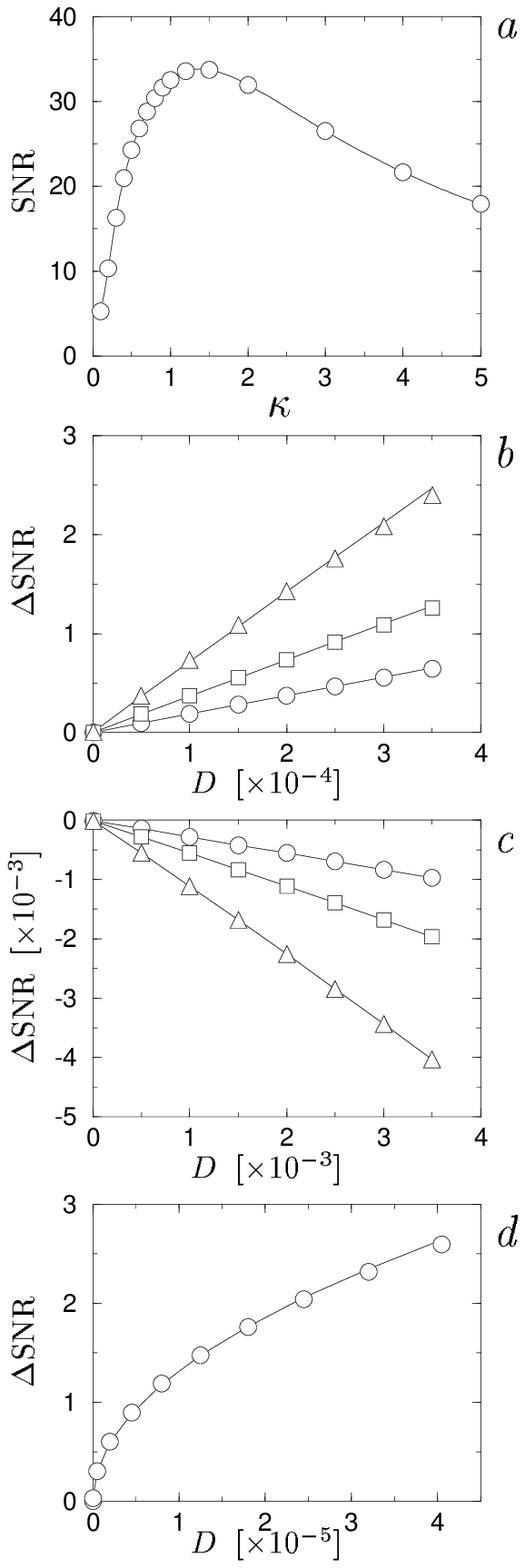}
}
\caption[a]{\label{fig1}
(a) SNR from the linearized SH equation as a function of $\kappa$
for $\alpha=1$, $q=1$, $\omega_0/2\pi=0.195$, and $g=1$.
(b)  $\Delta  {\rm  SNR} = {\rm  SNR}-{\rm  SNR}(D=0)$  from  Eq.
(\ref{SH})  as  a  function  of  $D$  for  $\kappa=0.3$,   $q=1$,
$\omega_0/2\pi=0.195$, and $g=0.5$ (circles), $g=1$ (squares) and
$g=2$  (triangles).  The points  have been fitted by a power law,
obtaining  an  exponent  equal to $0.99$ for the three  values of
$g$.
(c) $\Delta {\rm SNR}$ as in case b) but for $\kappa=3$.  In this
case the exponents of the power law are $0.99$, $1.00$, $1.02$.
(d) $\Delta {\rm SNR}$ from Eq.  (\ref{SH})  when  replacing  the
nonlinear  term  by  $g\psi|\psi|$,  as a  function  of  $D$  for
$\kappa=0.3$,  $q=1$, and  $\omega_0/2\pi=0.195$,  and $g=1$.  In
this case the scaling exponent is $0.50$.
In all figures the size of the system is $32$.
}
\end{figure}

To verify these results we have integrated the previous equations
by discretizing  them on a mesh  \cite{press} and then by using a
standard   method   for   stochastic    differential    equations
\cite{kloeden}.  For  the  sake  of  simplicity,  we  will  first
consider the one-dimensional  case, although the previous scaling
laws hold independently on the  dimensionality of the system.  In
Fig.  \ref{fig1}(a)  we have  depicted  the SNR as a function  of
$\kappa$ for particular values of the remaining  parameters.  One
can see that this quantity has a maximum at $\kappa_{max}$.  As a
consequence,  in view of Eq.  (\ref{SNR2})  the SNR may  increase
($\kappa<\kappa_{max}$) or decrease  ($\kappa>\kappa_{max}$) with
$D$.  In   Figs.  \ref{fig1}(b)   and   \ref{fig1}(c)   we   have
represented    the    SNR   as   a    function    of   $D$    for
$\kappa<\kappa_{max}$  and  $\kappa>\kappa_{max}$,  respectively.
These figures  corroborate the dependence of the SNR on the noise
level  and  on  the  parameter  $g$  in  Eq.  (\ref{SNR2}).  This
scaling  argument is robust upon varying the nonlinear term.  For
example, if one replaces the term $g\psi^3$ by $g|\psi|\psi$  one
obtains a SNR that  increases  as  $\sqrt{D}$  instead of as $D$.
This result is shown in Fig \ref{fig1}(d).  In this regard, other
nonlinearities have also been successfully tested.

\begin{figure}[t]
\centerline{
\epsfxsize=6cm 
\epsffile{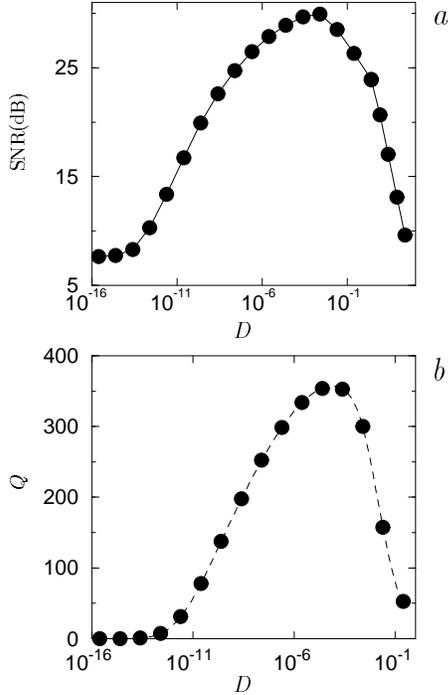}
}
\caption[b]{\label{fig2}
(a) SNR and (b) $Q$ as  functions  of the  noise  level  $D$  for
$\alpha=1$,  $\kappa=0.1$,  $g=1$, 
$q=1$,  and  $\omega_0/2\pi=0.024$.  The system
size is $32$.
}
\end{figure}

\begin{figure}[t]
\centerline{
\epsfxsize=8cm 
\epsffile{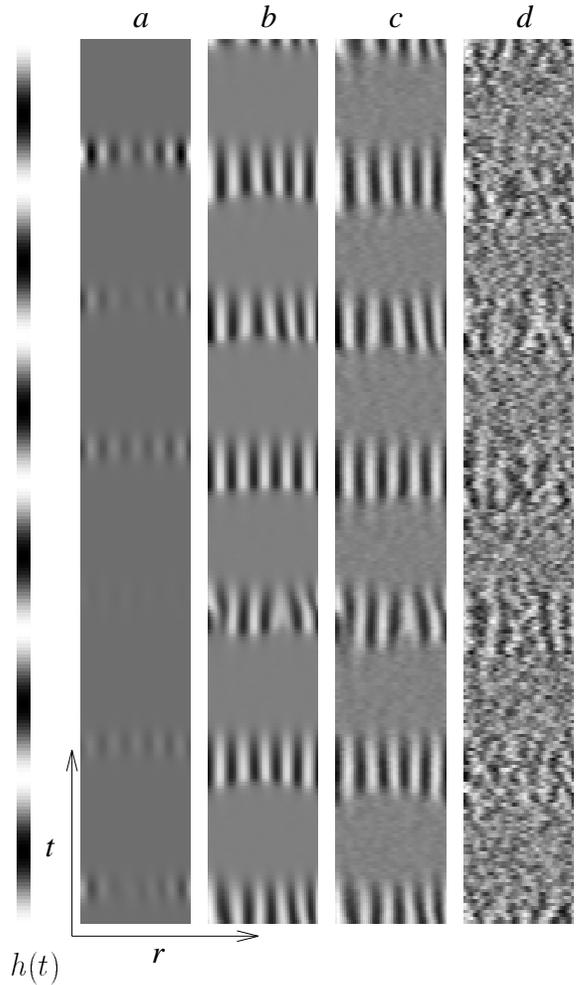}
}
\caption[c]{\label{fig3}
Spatiotemporal  evolution of the one-dimensional SH equation with
periodic   boundary   conditions   for  the  noise   levels   (a)
$D=0.5\times10^{-15}$,      (b)     $D=0.5\times10^{-4}$,     (c)
$D=0.5\times10^{-2}$,   and  (d)   $D=0.5$.  The  values  of  the
remaining parameters are $\alpha=1$,  $\kappa=0.1$, $g=1$, $q=1$,
and  $\omega_0/2\pi=0.024$.  The  system  size is  $32$.  We have
also  indicated  the  time  evolution  of the  control  parameter
$h(t)$.  For each  pattern,  black  and  white  colors  stand for
minimum and maximum values, respectively.
}
\end{figure}

The  previous   analysis  has  shown  the  possibility   for  the
appearance  of SR in the SH equation.  In Fig.  \ref{fig2}(a)  we
have plotted the SNR for $J(t)$ as a function of $D$, observing a
maximum at a nonzero noise level.  It is worth  pointing out that
an optimized  amount of noise  increases the SNR up to 20 dB.  As
far as the spatial  effects are  concerned,  the sharpness of the
peak of $S(k)$ can be analyzed by considering its height over its
variance,  which will be denoted by $Q$.  This magnitude has also
a maximum [Fig.  \ref{fig2}(b)] which is due to the fact that for
low noise  level $Q$ scales  with $D$ in the same way as  $S(k)$,
whereas  for high  noise  level the noise  destroys  the  spatial
structure.  An  example of the  spatiotemporal  evolution  of the
system is  depicted  in Fig.  \ref{fig3}.  For  sufficiently  low
noise  level, the system  exhibits  neither  spatial nor temporal
structures.  For  intermediate  values of the  noise,  the system
shows a spatial  pattern and a coherent  response to the periodic
variations of $h(t)$.  These  patterns are  destroyed  for higher
values of $D$.  In this  case,  noise  induces  periodic  spatial
patterns   and  ordered   temporal   behavior,   thus  playing  a
constructive role in both space and time.

\begin{figure}[t]
\centerline{
\epsfxsize=8cm 
\epsffile{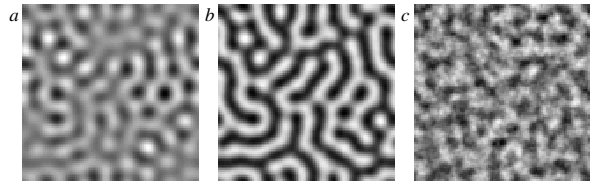}
}
\caption[d]{\label{fig4}
Representation  of the field  $\psi$ for the  two-dimensional  SH
equation with periodic  boundary  conditions.  The noise level is
(a)  $D=0.25\times10^{-10}$,  (b)  $D=0.25\times10^{-1}$, and (c)
$D=2.5$.  The remaining values of the parameters in all cases are
$\alpha=1$,       $\kappa=0.1$,       $g=1$,      $q=1$,      and
$\omega_0/2\pi=0.012$.  The system size is $50\times50$.
}
\end{figure}

The characteristics  observed for the one-dimensional SH equation
also hold in two  dimensions.  To illustrate  the effect of noise
in the  two-dimensional  case, we have plotted three patterns for
different  values of $D$ in Fig.  \ref{fig4}.  It  becomes  clear
that there exists an optimum  noise level in order to observe the
typical convective rolls.

Our results can also be applied to a great  variety of systems in
the  vicinity of a  bifurcation  point.  We note in this  context
that the form of the  scaling  laws does not depend on the way in
which  the nabla  operator  enters  the  linearized  equation.  A
remarkable  example, which enables one to predict the presence of
STSR by only  considering  scaling  arguments, is the generalized
Ginzburg-Landau  equation  \cite{van,cross}.  In  addition to the
cubic term discussed previously, this equation presents a quintic
one responsible for saturation  effects.  Since in this situation
the  coefficient  of the cubic term may be positive or  negative,
there will be a case in which the SNR is an  increasing  function
of the noise level giving rise to the appearance of STSR.

In summary, for the first time we have shown the presence of STSR
arising  when  the  system   undergoes  a  bifurcation,  a  usual
mechanism for pattern  formation.  One aspect to be emphasized is
the fact that in spatial nonlinear systems, the presence of noise
in  combination  with a periodic  signal may give rise to ordered
spatiotemporal structures which are not present in the absence of
noise.  We have also explicitly shown the presence of STSR in the
SH equation,  due to its great  importance  as a model of pattern
formation.  However, our results can be straightforwardly applied
to a wide variety of systems,  provided  they  exhibit the common
characteristics  of this  bifurcation.  These findings  therefore
open up new perspectives  about the general  consideration of the
phenomenon of SR to the field of pattern-forming systems.

This work was supported by DGICYT of the Spanish Government under
Grant No.  PB92-0859.  J.M.G.V.  wishes to thank  Generalitat  de
Catalunya for financial support.

\end{document}